\input harvmac
\def \h {\hat}

\def\const {{\rm const}}
\def \s {\sigma}
\def \p {\phi}
\def \h {\hat }
\def \ha {\half}
\def \ov {\over}

\def \a {\alpha}
\def \lr { \lref}
\def\ep{\epsilon}

\def\const {{\rm const}} \def\m{\mu}\def\n {\nu}\def\l
{\lambda}
\def \del {\partial}

\def \ha{{\textstyle{1\over 2}}}
\def \sixth {{\textstyle{1\over 6}}}
\def \eight {{\textstyle{1\over 8}}}

\def \D {\Delta}
\def \a {\alpha}

\def \zeta {\zeta}
\def \s {\sigma}
\def \p {\phi}
\def \m {\mu}
\def \n {\nu}

\def \t {\tau}
\def \td {\tilde }

\def \sm {$\s$-model\ }

\def \inv {^{-1}}
\def \ov {\over }

\def \third {{\textstyle{1\over 3}}}

\def \d {\delta} 
\def \F {\td F} 
\def   \td {\tilde }
\def \k {\kappa}
\def \A {\td A}
\def \L {\Lambda}
\def \F {{\cal F}} 
\def \db {$D$-brane }\def \dbs {$D$-branes }

\def\np {  Nucl. Phys. }
\def \pl { Phys. Lett. }

\def \prl { Phys. Rev. Lett. }
\def \pr  { Phys. Rev. }
\def \cqg { Class. Quant. Grav.}

\def \ijmp { Int. J. Mod. Phys. }

\baselineskip8pt
\Title{\vbox
{\baselineskip 6pt{\hbox{  }}{\hbox
{Imperial/TP/95-96/26  }}{\hbox{hep-th/9602064}} {\hbox{  }}} }
{\vbox{\centerline {Self-duality of Born-Infeld action  }
 \centerline { and Dirichlet 3-brane of type IIB superstring theory }
 }}
\medskip
\centerline{   A.A. Tseytlin\footnote{$^{\star}$}{\baselineskip8pt
e-mail address: tseytlin@ic.ac.uk}\footnote{$^{\dagger}$}{\baselineskip8pt
On leave  from Lebedev  Physics
Institute, Moscow.} }

\smallskip\smallskip
\centerline {\it  Theoretical Physics Group, Blackett Laboratory,}
\smallskip

\centerline {\it  Imperial College,  London SW7 2BZ, U.K. }
\bigskip\bigskip
\centerline {Abstract}
\medskip
\baselineskip10pt
\noindent
\medskip
$D$-brane actions depend on a world-volume abelian vector field
and are described by Born-Infeld-type actions. We consider the 
vector field duality transformations of these actions. Like the 
usual 2d scalar duality rotations of isometric string coordinates
imply target space $T$-duality, this vector duality is intimately
connected with $SL(2,Z)$-symmetry of type IIB superstring theory. 
We find that in parallel with generalised 4-dimensional Born-Infeld action, the action of 3-brane of type IIB theory is $SL(2,Z)$ self-dual. 
This indicates that 3-brane should play a special role in type IIB 
theory and also suggests a possibility of its 12-dimensional reformulation.

\Date {February 1996}

\noblackbox
\baselineskip 16pt plus 2pt minus 2pt

\lr \chap {A.H. Chamseddine, \np B185 (1981) 403;
E. Bergshoeff, M. de Roo, B. de Wit and P. van Nieuwenhuizen, \np B195 (1982)
97;
G.F. Chaplin and N.S. Manton, \pl B120 (1983) 105.}

\lr \ft { E.S. Fradkin  and A.A. Tseytlin, \pl B160 (1985) 69.  }

\lr \ftt { E.S. Fradkin  and A.A. Tseytlin, \pl B158 (1985) 316;  
 A.A. Tseytlin,  \pl B208 (1988) 221. }
\lr \tsee {Tseytlin }

\lr \gs{ M.B. Green  and J.H. Schwarz, \pl B149 (1984) 117;
\pl B151 (1985) 21; \np B255 (1985) 93. }

\lr \callan {C.G. Callan, C. Lovelace, C.R. Nappi and S.A. Yost, 
\np B308 (1988) 221.
}
\lr \abo{
A. Abouelsaood, C. Callan, C. Nappi and S. Yost,  \np B280 (1987) 599.}
\lr \bbe{
E. Bergshoeff, E. Sezgin, C.N. Pope  and P.K. Townsend, \pl B188 (1987) 70.}

\lr \duf { M.J. Duff, P.S. Howe, T. Inami and K.S. Stelle, 
\pl B191 (1987) 70. }

\lr \pol {J. Polchinski, ``Dirichlet-branes and Ramond-Ramond charges", NSF-ITP-95-122, hep-th/9510017.}

\lr\poly{D. Polyakov,  ``R-R - dilaton interaction in type II superstring",
RU-95-85,  hep-th/9512028.}
\lr \hulb{E. Bergshoeff, C.  Hull  and T. Ortin, \np  B451 (1995) 547, hep-th/9504081.}
\lr\gups{Gupser}
\lr \seza {M. Huq and M.A. Namazie, \cqg 2 (1985) 293, 597 (E); 
F. Giani and M. Pernici, \pr D30 (1984) 325;
I.C. Campbell and P.C. West, \np B243 (1984) 112;
S.J. Gates, Jr., J. Carr and R. Oerter, \pl  B189 (1987) 68.}
\lr \sezb{
J.H. Schwarz, \np B226 (1983) 269;
P.S. Howe and P.C. West, \np B238 (1984) 181.}

\lr \gs{ M.B. Green and J.H. Schwarz, \pl B136 (1984) 376; \np B243 (1984) 285.}

\lr \wiit{E. Witten, \np B266 (1986) 245; 
J.J. Atick, A. Dhar and B. Ratra, \pl B169 (1986) 54;
R. Kallosh, Phys. Scr. T15 (1987) 118;
M.T. Grisaru and D. Zanon, }

\lr \berg{ E. Bergshoeff, E. Sezgin and P.K. Townsend, \pl B189 
(1987) 75; Ann. of Phys.  185 (1988) 330. }

\lr \pkt {P.K. Townsend, \pl B350 (1995) 184.}
\lr \cree{E. Cremmer, B. Julia and J. Scherk, \pl B76 (1978) 409.}
\lr \ber {N. Berkovits and W. Siegel, ``Superspace effective actions for 4D compactifications of heterotic and type II superstrings", 
IFUSP-P-1180, ITP-SB-95-41,
hep-th/9510106.}
\lr \canon {A. Giveon, E. Rabinovici and G. Veneziano, \np B332 (1989)167;
K. Meissner and G. Veneziano, \pl B267 (1991) 33;
E. \' Alvarez, L. \' Alvarez-Gaum\' e and Y. Lozano, \pl B336 (1994) 183. }
\gdef \jnl#1, #2, #3, 1#4#5#6{ { #1~}{ #2} (1#4#5#6) #3}

\lr \cala{C. Callan, D. Friedan, E. Martinec and  M. Perry,  \np B262 (1985)
593.}

\lr \tsd{ A.A. Tseytlin,  \pl B242 (1990) 163;  \np B350 (1991) 395.}
\lr \sss {J.H. Schwarz and A. Sen,   \np B411 (1994) 35; \pl B312 (1993) 105.}
\lr \shr {E. Schr\"odinger, Proc. Roy. Soc.  A150 (1935) 465.}
\lr \tze {H.C. Tze, Nuov.Cim. 22A (1974) 507.}

\lr \gir {G.W. Gibbons and D.A. Rasheed, 
\np B454 (1995) 185, hep-th/9506035.}
\lr \ght {G.W. Gibbons, G.T. Horowitz and P.K. Townsend, \cqg 12 (1995) 297,
hep-th/9410073.}
\lr \girr {G.W. Gibbons and D.A. Rasheed, 
 \pl B365 (1996) 46, 
 hep-th/9509141.}
\lr \ft { E.S. Fradkin  and A.A. Tseytlin, \pl B160 (1985) 69.  }
\lr \frt{ E.S. Fradkin and A.A. Tseytlin, \pl {B163  }
(1985) 123;
R.R. Metsaev and A.A. Tseytlin, \np B298 (1988) 109;
O.D. Andreev and A.A. Tseytlin, \jnl \np, B311, 205, 1988.
}
\lr \mrt {R.R. Metsaev, M.A. Rahmanov and A.A. Tseytlin, \pl B193 (1987) 207;
A.A. Tseytlin, \pl B202 (1988) 81.}
\lr \andr{O.D. Andreev and A.A. Tseytlin, \pl B207 (1988) 157; 
 \np B311 (1988) 205.}
\lr \green{ M.B. Green, \pl B329 (1994) 435.}

\lr \dlp{
J. Dai, R.G. Leigh and J. Polchinski,
Mod. Phys. Lett. {A4} (1989) 2073.}
\lr \lei{
R.G. Leigh, Mod. Phys. Lett. {A4} (1989) 2767.}
\lr \pol{J. Polchinski, \prl  75  (1995)  4724,  hep-th/9510017.}

\lr\witt{ E. Witten, IASSNS-HEP-95-83, hep-th/9510135. }
\lr\bach{ C. Bachas,  NSF-ITP/95-144, hep-th/9511043.}
\lr\kle{C.G. Callan and I.R. Klebanov, 
 PUPT-1578, hep-th/9511173.}
\lr \doug{M. Douglas, RU-95-92, 
hep-th/9512077.}
\lr \schm{
C. Schmidhuber,  PUPT-1585, hep-th/9601003. }
\lr\gup{I.R. Klebanov and L. Thorlacius,  PUPT-1574, hep-th/9510200;
S.S.~Gubser, A.~Hashimoto, I.R.~Klebanov and J.M.~Maldacena, 
PUPT-1586, 
hep-th/9601057. }
\lr \pkt{P.K. Townsend, DAMTP-R/95/59,   hep-th/9512062. } 
\lr \blt{E. Bergshoeff, L.A.J. London and P.K. Townsend, Class. Quantum 
Grav. {9} (1992) 2545.}
\lr \deal{S.P. de Alwis and K. Sato, COLO-HEP-368, hep-th/9601167.}
\lr \jon {J.H. Schwarz,  \pl B360 (1995) 13; B364 (1995) 252 (E), 
hep-th/9508143, hep-th/9509148; CALT-68-2025, hep-th/9510086.}
 
\lr \ferr {S. Cecotti and  S. Ferrara, 
    \pl B187 (1987) 335.}
\lr \vaf {C. Vafa,  HUTP-96/A004, hep-th/9602022.}
\lr\tow {P.K.  Townsend p-brane democracy hep-th/9507048}
\lr \hull {C.M. Hull, QMW-95-50, hep-th/9512181.}
\lr \hut { C.M. Hull  and P.K. Townsend,  \np B438 (1995) 109. }
\lr \www {E. Witten,  \np B443 (1995) 85. } 
\lr \hos {G. Horowitz and A. Strominger, \np  B360 (1991) 197. }
\lr \dul { M.J. Duff and  X. Lu,  \pl  B273  (1991) 409.}
\lr \dkl  {M.J. Duff, R.  Khuri and  X. Lu,   Phys.  Repts.  259 (1995) 213.}
\lr \spec{M.J.  Duff, unpublished; P.K.  Townsend, unpublished }
\lr \hulb{E. Bergshoeff, C.M.  Hull  and T. Ort\'in, \np  B451 (1995) 547, hep-th/9504081.}
\lr \bbb{E. Bergshoeff,  H.J. Boonstra  and T. Ort\'in, 
UG-7/95,  hep-th/9508091.}
\lr \dub {M. Blencowe and M.J.  Duff, \np B310 (1988) 387.}

\lr \adem { M.~Ademollo, L.~Brink, A.~D'Adda, R.~D'Auria, E.~Napolitano,
        S.~Sciuto, E.~Del Guido, P.~Di Vecchia, S.~Ferrara, F.~Gliozzi,
        R.~Musto, R.~Pettorini and J.H.~Schwarz, 
\np B111 (1976) 77;  \np B114 (1976) 297. }
\lr \bri { L.~Brink and J.H.~Schwarz, \np B121 (1977) 285.  }
\lr \fraa {  E.S.~Fradkin and A.A.~Tseytlin, \pl B106 (1981) 63.}
\lr \adli {A.~D'Adda and F.~Lizzi, \pl 191 (1987) 85;
S. Mathur and S. Mukhi, \pr D36 (1987) 465; H. Ooguri and C. Vafa, \np B361 (1991) 469. }
\lr \marcus {N. Marcus, `A tour through N=2 strings', hep-th/9211059. }
\lr \tse { A.A. Tseytlin, \pl B367 (1996) 84,    hep-th/9510173 }
\lr \lii{M. Li, BROWN-HET-1020, 
 hep-th/9510161. } 

\lr \ttt{A.A. Tseytlin,  \pl B251 (1990) 530.}

\lr \ftt { E.S. Fradkin  and A.A. Tseytlin, \pl B158 (1985) 316; \pl B160 (1985) 69.} 
 \lr \aat{ A.A. Tseytlin,  \pl B208 (1988) 221;
\ijmp A4 (1989) 1257. }

\lr \ft { E.S. Fradkin  and A.A. Tseytlin, \pl B160 (1985) 69.  }
\lr \frt{ E.S. Fradkin and A.A. Tseytlin, \pl {B163}
(1985) 123.}
\lr \mrt {R.R. Metsaev, M.A. Rahmanov and A.A. Tseytlin, \pl B193 (1987) 207;
A.A. Tseytlin, \pl B202 (1988) 81.}

\lr \mett{R.R. Metsaev and A.A. Tseytlin, \np B298 (1988) 109;
A.A. Tseytlin, \ijmp  A5 (1990) 589.}
\lr \sezb{
J.H. Schwarz, \np B226 (1983) 269;
P.S. Howe and P.C. West, \np B238 (1984) 181.}
\lr\mark{ 
N. Marcus and  J.H. Schwarz,  \pl B115 (1982) 111.  } 
\lr \flor{R. Floreanini and  R. Jackiw,  \prl 59 (1987) 1873;
M. Henneaux and C. Teitelboim, \pl B206 (1988) 650. }
\def \S {\td S} 
\lr \hora{ P.  Horava, \pl B231 (1989) 251.}
\lr \bii {M. Born and L. Infeld, Proc. Roy. Soc. A144 (1934) 425.}
\lr \kut{D. Kutasov and E. Martinec, EFI-96-04, hep-th/9602049.}
\lr \duff{M.J. Duff, J.T. Liu and  R. Minasian,  
 \np B452 (1995) 261;  hep-th/9506126. }
\lr \duuf{ M.J. Duff  and J.X. Lu,  \np B390 (1993) 276,  hep-th/9207060.} 

\lr \ova{ H. Ooguri and C. Vafa, \np B361 (1991) 469;
\np B367 (1991) 83. }

\lr\jjj{J. Polchinski, S. Chaudhuri and C.V. Johnson, NSF-ITP-96-003, hep-th/9602052.}
\lr\jj{M.B. Green, DAMTP/96, hep-th/9602061.}
\lr \gri{ M.B. Green, \np B293 (1987) 593.}
\lr \old {B.M. Barbashov and N.A. Chernikov, Sov. Phys. JETP. 23 (1966) 861.}

\newsec{Introduction}
Type II superstring theories have supersymmetric $p$-brane solitonic
solutions supported by Ramond-Ramond (R-R) sources \refs{\hos,\dul, \dkl}.
These solitons  have  alternative  description 
in terms of open strings with Dirichlet boundary conditions.
$D$-branes  \refs{\dlp,\lei,\green}
provide an important tool \pol\ 
to study non-perturbative 
properties \refs{\hut,\www}  of superstring theories  
(see, e.g.,  \refs{\witt,\lii,\gup,\bach,\kle,\doug,\jjj,\jj}). 
They  are described by an effective   Dirac-Born-Infeld (DBI) 
action \lei\ (supplemented with extra couplings to R-R fields
\refs{\lii,\doug,\pkt,\schm,\deal})  which is 
closely connected to  the Born-Infeld (BI) type  effective action
of open string theory  \refs{\frt,\abo,\bbe,\mrt,\andr,\callan}.

The (bosonic part of) the DBI action of a $D$-p-brane is given by
the $d=p+1$ dimensional world-volume integral 
and depends  on  
the  embedding coordinates $X^\m$ and  the 
world-volume  vector vield $A_m$.  
Our aim below will be to consider in detail  the case of the 
$D$-3-brane  of  type IIB theory which (like the 2-brane 
in type IIA  case)  should play a special role 
in this  theory. 
The corresponding  action  can be interpreted as 
a generalisation of $d=4$ BI action coupled to a special 
 background metric, dilaton, axion, etc.
Using the remarkable fact 
that,   like Maxwell action, the  $d=4$ BI action  is invariant \refs{\shr,\tze,\gir} under 
the semiclassical vector  
duality transformation  ($A_m \to \td A_m$),   we shall demonstrate 
that the $D$-3-brane action (combined with type IIB effective action) is invariant under the 
$SL(2,Z)$ symmetry of type IIB theory \refs{\sezb,\hut,\www}.\foot{This confirms the conjecture made in \duff.
The $SL(2,R)$ duality invariance of  equations of motion following from 
a similar BI-dilaton-axion action 
was  first proved  in \girr\  but a connection of their  action 
to string theory was unclear. The  $D$-3-brane action
invariant under $SL(2,R)$ symmetry of type IIB theory  
provides the   proper context for  the application
of  the result of   \girr.}
This  is consistent with  expectations  \refs{\jon,\witt} 
that while there are $SL(2,Z)$ multiplets of  type IIB string  solutions, 
 the self-dual 3-brane solution \refs{\hos,\dul} should be unique.

As in the case of  the relation between the scalar-scalar  world-sheet duality
and $T$-duality of string theory,  we suggest that  the 
world-volume vector duality of $D$-brane actions is  a key to
 understanding 
of the $SL(2,Z)$ symmetry  of type IIB theory. 
The presence  in the $D$-3-brane action of a propagating world-volume vector field with 2  physical degrees of freedom  (and 
analogy with $D$-2-brane  related \refs{\pkt,\schm} to the 11-dimensional 
supermembrane \berg) 
points towards  possible existence of a (non-Lorentz-covariant) 
12-dimensional $D$-3-brane  formulation of type IIB theory.

As a preparation for a discussion of 
 $D$-brane world-volume vector field duality transformations 
   we shall  first study  (in Section 2)   the  semiclassical duality 
transformations of Born-Infeld actions 
 in different dimensions. 

The $D$-brane actions in type II theories 
may be found   by computing the superstring 
partition function  in the presence of a  $D$-brane 
 hypersurface which
can be probed by 
 virtual closed strings only through 
virtual open strings with ends attached to the surface \refs{\dlp}.
In Section 3  we shall present a  direct path integral derivation 
of the DBI action  \lei\  which explains its relation to $D=10$ BI action.
We shall then include the couplings to the background fields 
of R-R sector \refs{\doug,\schm}
and   obtain   the explicit form 
 (of leading terms  in derivative expansions) of 
$D$-p-brane actions for $p=1,2,3$. 
 
The duality transformations of these actions will be considered 
in Section 4. The cases of $p=1,2$ were  previously  discussed 
in \refs{\schm,\deal} but our approach and interpretation 
is somewhat different. 
We shall find that  while the action of  $D$-string   
 transforms covariantly  under the $SL(2,R)$ duality of type IIB theory, 
the $D$-3-brane  action is  `self-dual', i.e. is 
mapped into itself provided one also
duality-transforms  the $d=4$ world-volume 
 vector field.  

Some  conclusions  which follow from the  analysis 
of $D$-brane actions will be presented in Section 5.
In particular, we shall suggest that like the $D$-2-brane
in type IIA theory, the   $D$-3-brane  plays  a  special 
role in type IIB theory.  We shall  also argue that  the    
analogy with 2-brane case  indicates   that the  dimension 
`associated' with $D$-3-branes is 12
(different remarks on possible  12-dimensional 
connection of type IIB theory appeared  
in  \refs{\dub,\hull,\vaf}).

\newsec{Duality transformations  of Born-Infeld actions}
Before turning to  $D$-brane  actions let us discuss 
semiclassical duals of similar  Born-Infeld actions in different dimensions. 

Dualities between various gaussian fields (scalar-scalar in $d=2$, scalar-vector in $d=3$,  vector-vector in $d=4$, etc.) 
are,  in general,  true at the quantum (path integral level.
Restricting  consideration to   semiclassical level
duality transformations can be applied   to more general 
  non-gaussian  actions depending only on field strengths.
Such  `semiclassical duality' is 
a relation between  the  two actions which lead to   dual sets of  
classical equations and have  the same  `on-shell' value.
Suppose  $L(F)$ is a Lagrangian that depends
on a $n$-form field in $d$ dimensions  only through 
 its  field strength.
Then  
$L'(F,\A ) = L(F) + i F d\A $   where 
$\A$ is a $\td n$-form field  ($\td n= d-n-2)$, 
is equivalent to $L(dA)$.
If $L$ is an {\it algebraic} function of $F$ 
(i.e. it  does not depend on  derivatives of $F$) 
then solving $\del L'/\del F=0$ 
 for $F$ and substituting the solution 
 into $L'$  will give a {\it local}
`dual' Lagrangian $\td L( d \td A)$.\foot{All such 
 dualities are  symmetries of equations of motion 
combined with Bianchi identities,  or maps between one action and a dual one (with equivalent sets of equations of motion).
To have duality as a  manifest symmetry of a (non-Lorentz-invariant)  action 
one is to double the number of  
field variables by introducing the `dual fields'
on the same footing as the original fields 
\green\refs{\tsd,\sss}.} 
  
In  the special  
case of $n=\td n= \ha (d-2)$ ($n=0$ in $d=2$, $n=1$ in $d=4$, 
$n=2$ in $d=6$, etc.) it  may happen that  for  some  
$L$ its  `semiclassical dual' 
$\td L$  is the same function of $d\A$ as 
$L$ is of $dA$.  
 Such $L$  can be called `self-dual'
in the sense that 
that duality maps $L(dA)$ into $L(d\A)$  (in general, 
up to an additional transformation 
of  `spectator'  background fields).  
The usual  gaussian  choice $L = F^2$ 
is of course of  that type.
A   remarkable  non-trivial example of a  semiclassically 
self-dual Lagrangian  is   the 
Born-Infeld  Lagrangian  in {\it four}  dimensions  \refs{\shr,\tze,\gir}.\foot{More general self-dual  $d=4$ vector Lagrangians $L(F)$
should satisfy a functional constraint which is a 
1-st order  Hamilton-Jacobi-type partial differential equation. 
It  has a  family of solutions parametrised by one function of one variable  \gir.
The  construction of non-gaussian  self-dual actions
for higher  $n=\td n= \ha (d-2)$ forms
(e.g. for $n=2$ in $d=6$) is,  in principle,
straightforward; in particular, there exists
 the direct   generalisation  of the BI action to $n>1$,  $d=2n +2$ \gir.}

Let us start with the  Born-Infeld action \bii\ 
in its simplest  flat-space form 
(generalisations to the presence of other fields 
and couplings are straightforward)
\eqn\bor{ S_d=  \int d^d x \sqrt {\det( \delta_{mn } + 
{ F}_{mn})} \ , \ \ \ \ \   F_{mn}  = \del_m A_n - \del_n A_m \ ,  }
where  we assume the $d$-dimensional space to have euclidean signature.
To perform the semiclassical duality transformation we  
add the Lagrange multiplier 
term  ($\L^{mn} =-\L^{nm}$)
\eqn\born{ S_d =  \int d^d x [ \sqrt {\det( \delta_{mn } + 
{ F}_{mn})}  +  \ha i \L^{mn} (F_{mn} - 2 \del_m A_n) ] \  
  , }
and first solve for  $A_m$, finding that  
\eqn\aa{ d=2:\  \L^{mn} = \ep^{mn} \L_0, \ 
\L_0=\const ;  \ \ \  \  d=3:\ \ \ \L^{mn} = \ep^{mnk} \del_k \A  ; }
$$
 d=4:\  \L^{mn} =  \ep^{mnkl} \del_k \A_l   ; \ \ \ 
d \geq 5: \   \L^{nk}  = {\textstyle {1\ov (d-3)!}} \ep^{nk m_1...m_{d-2}} \del_{m_1}\A_{m_2...m_{d-2}}
 .  $$
Then we solve for $F_{mn}$ 
 and get the dual action $\td S_d (\td F), \ \td F=d\A$.

The BI action is semiclassically equivalent to 
\eqn\bro{ S_d (F,V)=  
  \int d^d x\  [ \  \ha  V \det({ \delta_{mn } + { F}_{mn})}  + \ha  V\inv\ ]\ , }
where $V(x)$ is an auxiliary field. 
Using \bro\ leads to 
a slight simplification  in two and three dimensions
since   for  $d=2,3$ \ 
 $\det (\delta_{mn } + { F}_{mn}) = 1 + \ha F_{mn} F^{mn}$
so that 
  \bro\   is  quadratic 
in $F$ and  solving for $F$  in the analogue of \born\ 
 is straightforward. 
One ends up with 
\eqn\dua{ \S_{2,3} = \int d^d x [  \ha   V  +
 \ha   V\inv  ( 1 +  \ha  \L^{mn}\L_{mn}) ] \ , }
or, after  elimination of $V$, 
\eqn\dual{ \S_{2,3} =  \int d^d x 
\sqrt{  1 +  \ha  \L^{mn}\L_{mn} }= \int d^d x 
\sqrt{ \det( \delta^{mn } + { \L }^{mn})}  \ , 
}
$$ \S_2= \int d^2 x 
\sqrt{  1 +   \L_0^2 } 
\ , $$
$$
   \S_3= \int d^3 x 
\sqrt{  1 +   \del_m \A \del^m \A } = \int d^3 x 
\sqrt{ \det ( \delta_{mn}  +   \del_m \A \del_n \A)} \ . $$
Thus the  dual to BI action in $d=2$
is a  constant (or a  `cosmological term') \blt, 
while the dual in $d=3$ is a membrane-type action for a scalar
field.

Remarkably, the  second  form of  the dual action in  \dual\
is true also in $d=4,5$. 
A simple way to show this is to note that since 
the Lagrangian is  Lorentz ($O(d)$) invariant algebraic function of $F_{mn}$
 `integrating out' $F_{mn}$ can be done  (at each point $x$) in a 
 special Lorentz frame where 
$F_{mn}$  is block-diagonal with  eigen-values $f_1, ..., f_{[d/2]}$.
Then  ${\det( \delta_{mn } + 
{ F}_{mn})} =  {(1 +f_1^2)...(1+ f^2_{[d/2]})}$
and  we get from  \bro\ 
\eqn\rrr{ L=  \sqrt  {(1 +f_1^2)...(1+ f^2_{[d/2]})} + i \l_1f_1 +... + i \l_{[d/2]} f_{[d/2]}   \ , }
where $ \l_1 = \L^{12}, ...$\ .
In the case of $d=4,5$ \ 
${\det( \delta_{mn } + 
{ F}_{mn})} =  {(1 +f_1^2)(1+ f^2_{2})}$ so that 
 solving for $f_1,f_2$ and substituting back into the action 
gives
\eqn\rrre{ \td L=  \sqrt  {(1 +\l_1^2)(1+ \l^2_{2})}
=  \sqrt{ \det( \delta^{mn } + { \L }^{mn})} \ . }
Thus the  BI action in $d=4$  is `self-dual', i.e. 
\eqn\ddd{ \td S_4 = \int d^4 x \sqrt{\det (\delta_{mn}  + \td F_{mn})} \ , 
\ \ \   \td F_{mn} = \del_m \A_n - \del_n\A_m \ , }
 while  in 
$d=5$  it is dual to  the following antisymmetric 
tensor action 
\eqn\dddd{\td S_5 = \int d^5 x \sqrt{\det (\delta^{mn}  +  H^{*mn}) } \ , 
\ \ \  \ \   H^{*mn}  \equiv {\textstyle {1\ov 3! } }\ep^{mnkpq} H_{kpq} \ , 
\ \    H_{kpq} \equiv 3 \del_{[k}\A_{pq]} \ . }
The derivation  of the action dual to $d=4$ BI action can 
be generalised to the case of the presence  of an extra  
 axion  coupling term in \bor, 
\eqn\bord{ S_4=  \int d^4 x  \ [ \sqrt {\det( \delta_{mn } + 
e^{-{1\ov 2} \p} { F}_{mn})}   +   \eight i   \ep^{mnkl} C F_{mn} F_{kl} ] 
\ ,  }
where $C$ is a background   axion  field  and we have also introduced 
 a background 
 dilaton field $\p$  with $e^{{1\ov 2}\p}$ playing the role of an  effective 
gauge coupling constant.  
The analogue of \rrr\ then  is 
\eqn\rer{ L_4 =  \sqrt  {(1 + e^{-\p}f_1^2)(1+ e^{-\p}f^2_{2})} + i \l_1 f_1 + i \l_{2 } f_{2}   + i C f_1 f_2  \ .  }
Solving  for $f_1$ and  $f_2$  one finds 
 \eqn\rerw{  f_1 = -  i  \D\inv (  \sqrt{{\D +  \l_2^2 \ov \D  + \l_1^2 }}  \l_1 
  -i  e^\p C  \l_2 ) \  ,  } 
     $$ 
\td f_2 = -  i \D \inv (  \sqrt{{\D +  \l_1^2 \ov \D 
 +  \l_2^2 }}  \l_2   -    i  e^\p C  \l_1 ) \ ,  \ \  \ \ \D\equiv   e^{-\p} + e^\p  C^2 \ .  $$
The final expression for the dual  action  is  then   
$$ \td  S_4 
 =  \int d^4 x\  [ \sqrt  {(1 + e^{- \td \p}\l_1^2)(1+ e^{-\td \p}\l^2_{2})}    +  i  \td C \l_1 \l_2] $$
\eqn\rrw{ =  \   \int d^4 x  \ [ \sqrt {\det( \delta_{mn } + 
e^{-{1\ov 2} \td \p} {\td  F}_{mn})}   +  
 \eight i   \ep^{mnkl} \td C \td F_{mn} \td F_{kl} ] 
\ ,  }
where 
\eqn\dfd {  e^{- \td \p} = {1 \ov e^{-\p} +  e^{\p} C^2 } \ , \ \ 
\ \ \  \td C = - { C  e^{\p} \ov e^{-\p} +  e^{\p} C^2 } \ ,   }
 is the standard  inversion 
$\td \l= - 1/\l$, \ $\l\equiv  C + i e^{-\p}$. 
Since the action  \bord\ is also invariant under the shift
$C\to C + q,\  q=\const$,   the two transformations generate 
the full $SL(2,R)$  invariance  group 
\eqn\iin{ A_m \to \A_m \ , \ \ \ \   \l \to 
 \td \l= {p\l + q \ov r\l + s}\ , \ \ \ \l\equiv  C + i e^{-\p} \ . }
One can  also replace $\delta_{mn}$ in \bord\ 
by a general metric $g_{mn}$ 
which should be invariant under $SL(2,R)$.

We conclude that the generalised 
 BI action \bord\ is  $SL(2,R)$-self-dual, i.e. its form is invariant under the vector duality  accompanied   by the  $SL(2,R)$ 
transformation of the background fields $\p,C$.
  Equivalent  observation 
about  the $SL(2,R)$ invariance of equations of motion that follow from 
this  generalised axion-dilaton-BI action was made in \girr. 
As we shall see in Section 4.3,  a similar action  describes  the 
dynamics of $D$-3-brane of type IIB superstring theory. 


\newsec{ $D$-brane actions}
Below we shall  first  describe  a  direct  path integral  derivation 
of the D-brane actions explaining, in particular,  the relation between 
the BI action of \frt\  and DBI action of \lei. 
While  an  understanding  of a  fundamental string action
as a source added to the effective  string field theory action 
 needs  a non-perturbative  (`wormhole') resummation
of string loop expansion \ttt, the open string description \refs{\dlp,\pol}
of  \dbs  provides a simple perturbative recipe for deriving 
their actions.

Since  \dbs are described by  BI-type actions \lei\ 
we shall then (in Section 4) use the results of  Section 2 to find 
how these  actions change under vector duality transformations. 

\subsec{ Path integral derivation of $D$-brane  action }

\dbs represent  soliton configurations
in superstring theory which  have  R-R-charges \pol. 
Within  string perturbation theory they are    `composite'
objects which  have 
effective `thickness' 
of order  $\sqrt {\a'}$ \refs{\gup,\bach}.
As the effective  field theory action  $S_{eff}$ for massless 
string modes,   the effective action  $S_{D-brane}$ 
for a  \db  moving in a 
massless string  background  is  given by  
power series in $\a'$.\foot{In the case of a string 
($D=2$ effective theory) 
$\a'$-corrections should be trivial up to a field redefinition, i.e. 
should  affect only `propagator' terms.
At the same time, $D$-string is  expected to  become fundamental 
(`structureless') only  after  a resummation of string loops.}
The tree-level closed  string effective action 
$S_{eff}$  can be  represented in terms of  
(`renormalised') string partition 
on the  sphere \refs{\ftt,\aat}.
In the case of the open string theory the 
effective action is given by the  
partition function on
 the disc \refs{\frt,\mrt,\andr} (loop corrections can be 
 found 
by adding partition functions on higher genus surfaces \refs{\mett}).
 In the superstring case 
the leading term in the disc partition function 
is  finite (there is no  quadratic $SL(2,R)$ 
M\"obius volume infinity \andr) and is equal to the BI action.

Our  aim  is to find  the \db action  in a similar  way 
by evaluating  the string path integral in the presence of a $D$-brane.
In view of the open string  connection
 \dlp, the   \db effective action  (reconstructed in \lei\
 from the 
equations of motion 
obtained from  conformal invariance conditions  \refs{\cala,\abo})
can  be derived directly  by computing the
corresponding string partition 
function  on a disc. This is the  partition function
of virtual open strings with    mixed Dirichlet-Neumann boundary 
conditions (i.e. with ends attached to a hyperplane)
 propagating in a  condensate of massless  string modes. 
 The collective coordinates $X^i$ and internal vector 
$A_m$ degrees of freedom 
of the \db  are represented by the boundary  background 
couplings  \refs{\dlp,\lei}.
The combined  action of the string massless modes and \db source 
can be represented  as ($t$ is a logarithm of 2d cutoff)
\eqn\comb{ S_{eff}(G,B,\p;C)  + S_{D-brane}(X^i,A_m; G,B,\p;C)
= ({\del \ov \del t} Z_{sphere})_{t=1} + Z_{disc}  , }
\eqn\zz{ Z_{disc} = \int [dx d\psi] e^{-I}\ ,  \ \ \ \ 
I = I_M  + I_{\del M} \ , } 
\eqn\vol{ 
I_M= {1 \ov 4\pi \a'}  \int d^2 \xi\  [\ ( \sqrt g  g^{ab} G_{\m\n}+ i \ep^{ab} B_{\m\n})
 (x) \del_a x^\m \del_b x^\n + \a' R^{(2)} \p(x)\ ]+ ... \  , }
\eqn\bou{ 
I_{\del M}=  {1 \ov 2\pi \a'} \int d\tau \ [ \  i A_m(x)\del_\t x^m 
+ X_i(x) \del_{\perp} x^i + \a' e K \p (x)\ ] 
+ ... \ . } 
Here  $C$ denotes the  background  $r$-form fields from  the 
R-R sector  ($C_1,C_3$ in type IIA and  $C,C_2,C_4$ in type IIB theories) and dots stand for fermionic terms.\foot{This description
applies only to `electric' ($p\leq 3$) $D$-branes which we shall consider
in what follows. To generalise it to `magnetic' ones one presumably is to double
the number of the R-R fields  in the action (cf. \refs{\tsd,\sss}).} 
The string coordinates $x^\m$ ($\m=0,1,...,9$) 
are split into $x^m$ ($m=0,...,p$) and $x^i$ ($i=p+1,...,9$)
with the boundary conditions $\del_{\perp} x^m|_{\del M} =0, \ x^i|_{\del M}=0$ so that the boundary couplings depend only on $x^m$.
 An alternative  possibility is  to use 
the boundary condition $x^i|_{\del M}= X^i(x^m)|_{\del M}$  while  setting $X_i$-coupling
in \bou\  to zero \lei.
This has an advantage of   making obvious 
the interpretation of $X^i$ as the collective coordinates of \db but is  less natural from the point of view 
of computing the vacuum partition function  and  also  breaks 
the symmetry between $A_m$ and $X_i$ in $I_{\del M}$.
The two approaches are simply   related   to 
 leading order  of semiclassical expansion \lei:
assuming $x^i|_{\del M}= X^i(x^m)|_{\del M}$ and making a shift $x^i \to x^i + X^i|_{\del M }$
so that  the new $x^i$ is subject to $x^i|_{\del M}=0$
one finds that  $I_{\del M}$  gets $X_i = G_{ij} X^j + ...$\ .
 In general 
the transformation between  them is 
complicated and should involve  redefinitions
of background fields.

Choosing  the flat 2d metric 
(and ignoring the  dilaton couplings) we can 
formally re-write the combined action $I$ as 
\eqn\tra{ I  =  {1 \ov 4\pi \a'}  \int d^2 \xi \bigg((\d^{ab} G_{\m\n}+ i \ep^{ab} B_{\m\n})
 (x)\del_a x^\m \del_b x^\n }
$$  + \  \del_a [ i \ep^{ab} A_m (x) \del_b x^m 
+ X_i(x) \del^a x^i]+... \bigg) \ . $$
Let us  first  assume that the closed string couplings do not depend on $x^i$
(and are slowly varying in $x^m$).
Then to  compute  the  partition function \zz\
we  may  first  do  explicitly  the gaussian integration  over $x^i$.
Because of the  absence of the zero mode of $x^i$ 
this  will  not produce the standard  $\sqrt {\det G_{ij}}$ factor
(usually present in the covariant \sm  measure).
The only change  will be  in the couplings 
in the  action \tra\ which  will 
 now depend   on  $d=p+1$  coordinates $x^m$ with 
standard Neumann  boundary conditions. Expanding $x^m(\xi) =x_0^m + y^m(\xi)$
one can shift the $B(x_0)\del y\del y$ term to the boundary. Integrating first over  the values of $y^m$ at the internal points
of the disc and then  over their  boundary values one
 finds as in \refs{\frt,\mett} that to  the leading 
order in expansion in derivatives of fields ($d=p+1$) 
\eqn\zzz { Z_{disc} = c_1 \int d^d x_0  \ 
e^{-\p }  \sqrt { \det ( \h G_{mn} + \h B_{mn} + F_{mn}) } + ... \ ,  }
$$ \hat G_{mn} \equiv G_{\m\n} (x_0) \del_m X^\m \del_n X^\n\ ,  \ \ \ 
\hat B_{mn} \equiv B_{\m\n}(x_0)  \del_m X^\m \del_n X^\n\ , $$
$$
X^\m \equiv (x^m_0, G^{ij} X_j (x_0)
),  \ \ \ \del_m = {\del\ov \del x^m_0} , \ \ \  \  F_{mn} = \del_m A_n - \del_n A_m. $$
Equivalently,  we may compute $Z_{disc}$ in a more `10d-symmetric' way 
by  first  performing  the duality transformation \dlp\ 
$x^i \to \td x_i$  in \tra.  Then the boundary term takes 
the standard open string form with $A_\m=(A_m, X_i)$
so that the result is just the usual $D=10$ BI  action 
for the $O(9-p,9-p)$  dual background 
up to the factor  (which as usual can be absorbed into the dilaton)
accounting  for the fact that $\td x_i$ 
does not have zero mode part 
\eqn\zze{ Z_{disc} = c_1 \int d^d
 x_0\ e^{-\p}{1\ov \sqrt {\det(\td G_{ij} +  \td B_{ij})} }
  \sqrt { \det (  \td G_{\m\n} + \td  B_{\m\n} + F_{\m\n}) } + ... \ , }
$$  \td G_{ij} + \td  B_{ij} = (G_{ij} + B_{ij})^{-1} \ , 
\ \ \     \td G_{im} + \td  B_{im} = (G_{ij} + B_{ij})^{-1}
(G_{jm} +  B_{jm}),  .... \   , $$
\eqn\fff{ F_{mn} = \del_m A_n - \del_n A_m, \ \ \ F_{mi} = \del_m X_i\ , 
\ \ \   F_{ij} = 0  \ .  }
The two expressions \zze,\zzz\   agree  
 for constant $G,B$.

The  background fields  $G,B,\p$ in \zzz\ depend only on  $x^m_0$. 
When their dependence on $x^i$ is taken into account in
\tra\ one expects to find \zzz\
with  the fields depending on  $X^\m(x_0)= (x_0^m,\ X^i (x_0))$.
This indeed  is what happens  since one can re-introduce the zero mode for $x^i$ by the  shift $x^i \to x^i + X^i(x_0)$ which eliminates the  linear 
 boundary coupling term 
$X_i(x_0) \del_{\perp} x^i$. 

The resulting   $D$-p-brane  action   $S_{d} = Z_{disc}$ \  is the same as the DBI action of \lei\ (here $d=p+1$; in what follows we  omit the index $0$ on $x^m$)
\eqn\ddw{ S_d = \int d^{d} x \ e^{-\p} \sqrt{\det  ( \h G_{mn}  + \F_{mn}) } + ...\ , }
$$
 \F_{mn} = F_{mn} +  \hat B_{mn}\  , \ \ \ \  \hat G_{mn} +\h B_{mn}  =  (G_{\m\n} +B_{\m\n})(X) \del_m X^\m \del_n X^n\ .    $$
 It depends only on  transverse   \db  coordinates  
$X^i$ and on  $A_m$, i.e.   
corresponds
 to the static gauge $X^m(x)=x^m$. The number of physical modes 
 is the same for all  values of 
$d=p+1$: $10-d + d-2 =8$, 
where $d-2$ corresponds to a vector in $d$
dimensions.  In addition there are 8 fermionic modes as demanded
by unbroken  supersymmetry \pol\  (half of $N=2, D=10$ supersymmetry should be realised linearly and half -- in a non-linear way).
These on-shell degrees of freedom are  the  same as of $N=1,D=10$ Maxwell supermultiplet dimensionally reduced to $d$ dimensions \witt.

It is natural to  assume that the   above action
can be generalised
to  make  it   manifestly $D=10$ Lorentz and 
 world-volume  diffeomorphism 
invariant  by  relaxing the  static gauge condition, i.e. 
 treating  all $X^\m=(X^m,X^i)$  as  10 independent fields  in $d$
dimensions.  This will be assumed in what follows.\foot{It  
 is not  clear,  however,  
how  this can be done in  a manifestly supersymmetric way for  $d\geq 4$ 
(see also Section 5).}

\subsec{ Couplings to R-R background  fields}
In addition to the above couplings \ddw\  to NS-NS background fields $(G,B,\p)$, 
\db actions  contain couplings to  the 
 background fields  $C_r$ of R-R sector of type II theories.
The leading-order couplings  are  effectively   given by 
the fermionic zero mode ($\psi^m_0$) factor 
and can be computed systematically \refs{\lii,\doug, \schm,\deal} using the techniques  of  \refs{\callan}.
Since the   \sm action contains the boundary term 
$\F_{mn} \psi^m \psi^n$ 
plus couplings to R-R fields $C_r$ 
the resulting leading-order term \doug\
 has the  following
symbolic structure\foot{One  way to understand this 
expression is to note that the zero mode count on the disc implies that 
one of the R-R vertex operators should depend explicitly on 
the potential $C_r$ and not on its strength \jjj.}
  $\int  d^d x_0 d^d \psi_0  \sum C_r (\psi_0)^r  e^{\F\psi_0\psi_0}$.
In addition, for `magnetic' $D$-branes ($p\geq 3$)
the conformal invariance conditions imply that there should be 
 a source term in the  Bianchi identities for the field strengths $d C_r$  \refs{\pol,\doug,\schm}.

The explicit form of  these  coupling terms  
 thus  depends on particular $p$. 
For $p=d-1=1,2,3$ one finds 
\eqn\ddew{ S_2 = \int d^2 x\ [ \ e^{-\p} \sqrt {\det(\h 
 G_{mn}  + \F_{mn}) } + 
\ha i \ep^{mn} (\h C_{mn} +  C \F_{mn}) +...]\ , }
\eqn\dew{ S_3 = \int d^3 x \ [ \ e^{-\p} \sqrt{\det  (\h G_{mn}  + \F_{mn}) }  + 
 \ha i \ep^{kmn} (\third\h  C_{kmn} + \h  C_k \F_{mn}) +...]\ , }
\eqn\ded{ S_4 = \int d^4 x\ [\ e^{-\p} \sqrt {\det({\h G_{mn}  + \F_{mn}})} }
$$  + \ 
\eight i  \ep^{mnkl} (\third \h C_{mnkl} +    2\h  C_{mn}\F_{kl} 
+ C \F_{mn}\F_{kl}) +...]\ , $$
where  $\h C_m,\h C_{mnk}$  
and $ C,\h C_{mn},\h C_{mnkl}$ are projections of the 
  R-R fields of type  IIA and type  IIB theories ($\h C_{m} = C_{\m}(X)  \del_m X^\m $, etc.) 
and 
dots stand  for fermionic and  higher-order terms.\foot{The  constant $c_1$ 
 in front of the 
 actions \zzz,\ddw\  (proportional to the dilaton tadpole on the  disc) 
is  assumed to be 
 absorbed into $e^\p$ and normalisations of the fields $C_r$.}

 These  \db actions may   be combined  
according to \comb\ 
with the  effective action $S_{eff}$ 
for the  massless string  modes. 
For example, in the case of type IIB theory \sezb\  $S_{eff}$
is given by \refs{\hulb,\bbb}
\eqn\efec{  S_{eff \ IIB} 
= c_0 \int d^{10} x \sqrt{G} \bigg( e^{-2\p} [ R + 4 (\del \p)^2
- { \textstyle{3\ov 4}} (\del B_2)^2 ] 
} 
$$ -  \  \ha  (\del C)^2 
 - { \textstyle{3\ov 4}} (\del C_2  - C  \del B_2) ^2
- { \textstyle{5\ov 6}}  F^2(C_4) 
- { \textstyle{1\ov 48}} {\ep_{10}\ov \sqrt G} C_4 \del C_2 \del B_2 + ... \bigg) \ . $$
where $\del B_2= \del_{[\m} B_{\n\l]}$, etc., 
$F(C_4)= \del C_4 + {3\ov 4} (B_2 \del C_2 - C_2 \del B_2)$, 
and,  following \bbb, it is assumed   that the 
(conformally-invariant) self-duality constraint 
on $F(C_4)$  ($F= F^*$)
   is to be  added  at the level of equations of motion.
This action is invariant under
the $SL(2,R)$ symmetry \refs{\sezb,\hulb}
\eqn\sssl{ g_{\m\n} \to g_{\m\n} \ , \ \ \ 
\ \ \   \l \to {p\l + q \ov r\l + s } \ , }
$$
\ \ \  B_{\m\n} \to  s B_{\m\n} - r C_{\m\n}\ , \ \  \ \ \ 
C_{\m\n} \to  p C_{\m\n} - q B_{\m\n} \ , $$
\eqn\ssll{ 
   g_{\m\n} \equiv  e^{- {1 \ov 2} \p} G_{\m\n} \ , \ \ \ 
 \ \   \ \l\equiv  C + i e^{-\p} \ ,  }
which is  expected to be an exact  duality 
symmetry of type IIB superstring theory
 \refs{\hut,\www,\jon}.
A natural question is  whether  this  symmetry is present  in 
the {\it combined }  action of type IIB low-energy field  theory and 
a \db source.
As we shall  see  below,  the combined equations
of motion are indeed covariant under $SL(2,R)$
in the case of $D$-string and $D$-3-brane.
To demonstrate this it is necessary  to perform the 
world-volume vector  duality transformation
similar to the one discussed above for BI action.
 
\newsec{Duality transformations  of $D$-brane actions } 
Since the  
\db  actions \ddw--\ded\  depend  on  $A_m$ only through  $F_{mn}$ 
one can perform the semiclassical duality transformation as in  
the  case of the BI actions in Section 2, i.e.  by 
adding the  Lagrange multiplier 
term
 $ \ha i \L^{mn} (\F_{mn} - 2 \del_m A_n - \hat B_{mn})$ 
and eliminating  $\F_{mn}$ and $A_m$  from the action 
using their equations of motion.\foot{As was mentioned already, 
these  actions are low-energy effective actions (analogous, e.g., to Nambu-type  action for a cosmic string or domain wall) 
and may  be treated semiclassically. }
The resulting dual action has equivalent set of equations of motion
(with the roles of `dynamical  equations' and `Bianchi identities' interchanged).

\subsec{$D$-string}
Starting with \ddew\ we find  
that as in \born,\aa\ the $A_m$-equation  still  implies 
 $\L= \L_0=\const$
so that  \schm\ 
\eqn\ddew{ \S_2 = \int d^2 x\ [ \sqrt {e^{-2\p} + (\L_0 + C)^2} \ \sqrt {\det \h G_{mn}} + 
\ha i \ep^{mn} (\h C_{mn} - \L_0 \h B_{mn})]  \ . }
The dual to $D$-string action can thus be  interpreted 
as the standard fundamental  string action in  $SL(2,R)$ transformed 
metric and antisymmetric tensor background  
(cf.\sssl\ with $ p=0, \ q = -1 ,\  r=1, \ s= \L_0$,\ 
$\l \to -1/(\l +\L_0)$)
\eqn\ddeww{ \S_2 = \int d^2 x \ ( \sqrt {\det \hat G'_{mn}} + 
\ha i \ep^{mn} \hat B'_{mn} )  \ ,  }
\eqn\tre{ G'_{\m\n}  \equiv   \sqrt{  e^{-2\p} + (\L_0 + C)^2}\   G_{\m\n}, 
\ \ \ \    B'_{\m\n} \equiv   C_{\m\n} - \L_0 B_{\m\n}  \  ,   }
where $\hat G_{mn} = G_{\m\n} \del_m X^\m \del_n X^\n,$ etc.
This  action (and  its sum with \efec) 
is  thus covariant under the type IIB 
$SL(2,R)$ transformations \ssll\  of the background fields. 
Indeed, $ G'_{\m\n}$ can be written as
$ G'_{\m\n} = \sqrt{\Delta({\L_0})} g_{\m\n}, \ \ \ 
\Delta({\L_0}) =  e^{-\p} +  e^{\p} (\L_0 + C)^2 $, 
where $g_{\m\n}$ is  the $SL(2,R)$-invariant
Einstein-frame metric. Generic $SL(2,R)$ transformation
shifts the parameter $\L_0$ ($\L_0'=s\L_0 + q$) and also rescales the coefficient in front of the action (by factor $p+ r \L_0$).\foot{The parameter $\L_0$ takes discrete values 
at the quantum level \witt. In what follows we shall often 
not differentiate between $SL(2,Z)$ and $SL(2,R)$ transformations.} 
This duality covariance was not apparent before the duality 
transformation of $A_m$ (though it is, of course, 
  present also in  the equivalent set of equations that follow from \ddew).

Combined with the type IIB effective action \efec\ 
the $D$-string action 
plays the role of a source ($X^m=x^m, X^i=0$) 
  for  a set  of fundamental string solutions
of type IIB effective equations  constructed in \refs{\jon}.
These conclusions are consistent with 
the previous results about $SL(2,Z)$ covariant family of 
 type IIB strings \refs{\jon,\witt}  supporting  the  
$SL(2,Z)$  duality  symmetry of type IIB theory \refs{\hut,\www}.  
\subsec{$D$-2-brane }
In the case of $S_3$ \dew\  the Lagrange multiplier is 
$\L^{mn} = \ep^{mnk} \del_k \A $ 
(cf.\aa) 
and thus elimination of $\F_{mn}$  
(using, e.g., the representation \bro) gives 
the  dual action which is a generalisation 
of the expression in \dual\  (equivalent 
but less straightforward derivation of this action was given in \schm)
\eqn\dewq{ \S_3 = \int d^3 x \ \big[ e^{-\p} \sqrt {\det \{\h G_{mn}  + 
e^{2\p} (\del_m \A + \h C_m) (\del_n \A + \h C_n) \} }   } 
$$
+ \  \ha i \ep^{mnk} (\third \h C_{mnk} -  \h B_{mn}\del_k  \A) \big] \  . $$ 
This action  can be re-written in   the standard membrane action  form 
with one extra  scalar coordinate $\A$ 
\eqn\dewt{ \S_3 = \int d^3 x \ ( \sqrt{\det {\hat G'_{mn}}}  + 
  {\sixth}  i \ep^{mnl} \hat B'_{mnl} ) \ , }
$$  \hat G'_{mn} =  e^{-{2\ov 3 } \p} \hat G_{mn}  + 
e^{{4\ov 3}  \p}  (\del_m \A + \hat C_m) (\del_n \A + \hat C_n) , \ \ \  
\hat  B'_{mnl}= \hat C_{mnl} - 3\hat B_{[mn} \del_{l] }  \A   . $$
Dualising $A_m$ 
one thus  finds an  action which (for trivial 
background fields)  has
hidden global Lorentz symmetry $SO(1,10)$
of 11-dimensional theory as  previously suggested 
in \refs{\duuf,\duff}.
Because of  the same underlying  supersymmetry 
and field content (as implied by the 
 dimensional reduction 
relation between $D=11$ 
supergravity and  type IIA  low-energy effective action)
it is not actually  surprising to find that
the $D$-2-brane action is  dual to  the  direct 
  dimensional reduction (fields do not depend on $X^{11}\equiv \td A$)
of the 
  $D=11$ supermembrane  action coupled to $D=11$ supergravity background 
 \refs{\pkt,\schm}. 
The  degrees of freedom count and the 
 requirement of   supersymmetry  of a $D$-2-brane action
in a background of $N=2a$, $D=10$  supergravity  
uniquely  fixes its form  to be equivalent 
  to  that of the dimensionally 
reduced  
 $D=11$ supermembrane action \berg. This 
observation   can be used to  determine \pkt\  the fermionic terms in the 
$D$-membrane  action  \dewq\  by starting with the known $D=11$ supermembrane  action.

\subsec{$SL(2,R)$ self-duality of $D$-3-brane action}
Let us now turn to the case of our main interest
 -- the $D$-3-brane action  \ded\  
and perform the  world-sheet  vector duality transformation 
$A_m \to \A_m$ as in the case of the generalised $d=4$ BI action \bord,\rrw.  
Adding the Lagrange  multiplier term ($\L^{mn} =  \ep^{mnkp} \del_k \A_p$, cf. \aa) we find 
from \ded\ 
\eqn\dede{ S_4 = S_4' + S''_4, \ \  \ \ S''_4= \int d^4 x\    \eight i  \ep^{mnkl} (\third \h C_{mnkl} - 2 \h B_{mn} \td F_{kl}) + ... \ , }
\eqn\dde{ S_4' = \int d^4 x\ [e^{-\p} \sqrt {\det(\h G_{mn}  + \F_{mn})} 
  + 
\ha i {\td \L}^{mn} \F_{mn}  +  \eight i  \ep^{mnkl} C  \F_{mn}  \F_{kl }]  \ , }
$$
  {\td \L}^{mn} \equiv  \ep^{mnkl} (\td F_{kl} + \hat C_{kl})  \ ,  \ \ 
 \ \ \   \td F_{mn} = \del_m \A_n - \del_n \A_m \  .  $$
The remaining  problem of  eliminating   $\F$  from  $S'_4$
is solved as in the case of BI action \bord. 
Indeed,  the action \ded,\dde\ is closely related to \bord\ 
as can be seen by expressing it in terms of the 
$SL(2,R)$ invariant  $D=10$ Einstein-frame metric $g_{\m\n}$:  
$\h G_{mn}= e^{{1\ov 2}\p} \h g_{mn}$.
Then the dilaton dependence becomes the same as in \bord\ 
with $\h g_{mn}$ replacing $\d_{mn}$ there. 
As a result, the action dual to \ded\  is found to be\foot{We  assume that the action \ded,\dede\ contains a specific  `higher - order' term $\sim i \ep^{mnkl} $ $ \h B_{mn}\h C_{mn}$ as demanded by antisymmetric tensor gauge invariance.}
 \eqn\deqd{ \td S_4 = \int d^4 x\ [\ e^{-\td \p} \sqrt {\det(\h {\td G}_{mn}  + \td \F_{mn})} }
$$  + \ 
\eight i  \ep^{mnkl} (\third \h C_{mnkl} +    2\h {\td  C}_{mn}\td \F_{kl} 
+ {\td  C} \td \F_{mn}\td \F_{kl}) +...]\ ,  \ \ \ \ \ \td \F_{mn} = \td F_{mn}  + \h {\td B}_{mn} \ , $$
where  the transformed fields $\td \p, \td C$  
are  as  in \dfd\ and 
$$\td G_{\m\n} = e^{{1 \ov 2}( \td \p - \p) } G_{\m\n}\ , \ \ \  
 \td B_{\m\n} =  C_{\m\n}\ , \ \ \  \td C_{\m\n} = - B_{\m\n} \ . $$
As in the case of \rrw\ these  redefinitions of 
background fields  correspond to the  basic  $SL(2,R)$
duality transformation \ssll\ with $p=0,q=1, r=-1,s=0, \ \l \to - 1/\l$
and thus together  with trivial shifts of $C$ imply the 
full $SL(2,R)$ invariance of the action under 
$A_m\to \A_m$  combined  with the transformation  \ssll\ of background fields.

The duality invariance of the $D$-3-brane  action
is  thus closely related to the fact of self-duality 
of the  $d=4$  BI action 
   and  its $SL(2,R)$ generalisation  existing in  the case of specific 
dilaton-axion-BI system  first  noted    in  \girr.\foot{It was 
found in \girr\ 
  that  there  exists  the   unique 
generalisation   of the $d=4$  BI  action  to the case of  coupling   
to  the dilaton  $\p$ and axion  $C$ 
 whose equations of motion are invariant 
under  the $SL(2,R)$ duality transformations
generalising the vector duality transformations:
$S_{GR}=\int d^4 x \ [e^{-\p} \sqrt{ \det( e^{{1\ov 2} \p} g_{mn}+   F_{mn}) }
 +  \eight i  C \ep^{mnkl}  F_{mn}F_{kl}].$
This action  (which is the same as \bord\ with $\d_{mn} \to g_{mn}$) 
was combined in \girr\ with 
 the  standard  $d=4$  axion-dilaton action 
$ S_{eff}=-\int d^4 x \sqrt g [ R - \ha (\del \p)^2 - \ha e^{2\p} (\del C)^2].$
The result $S=S_{eff} + S_{GR}$ 
 is {\it not}, however,  the  effective  action 
that appears (upon  dimensional reduction to $D=4$)  in 
 type I superstring  theory 
 (or that may appear \tse\ in $SO(32)$ heterotic string theory 
as  suggested by its duality to type I theory in $D=10$):
 for $g_{mn}$ to be  the   Einstein-frame metric, 
its factor should be $e^{-2\p}$, not $ e^{{1\ov 2}\p}$ as in $S_{GR}$.
That remained a  puzzle in \refs{\gir,\girr}.
As we have seen  above, an  action 
closely related  to $S_{GR}$  does appear
in string theory -- as the action of $D$-3-brane 
of type IIB superstring theory.
In {\it ten}  (but not in four) dimensions
$G_{mn} = e^{{1\ov 2}\p} g_{mn}$
is indeed the transformation 
from the string frame to the $SL(2,R)$ invariant Einstein-frame metric.
 The analogue of the invariant   action  $S=S_{eff}+ S_{GR}$
  is  the sum of $D=10$ effective action  $S_{eff\ IIB}$ \efec\
and 3-brane action \ded\ and 
the $SL(2,R)$ symmetry in question is not  
the  usual $S$-duality of  $d=4$ effective  string  action  but
$SL(2,R)$ duality  of $D=10$ type IIB superstring theory. }

Thus we find that  in contrast to the case of 
$SL(2,R)$ {\it covariant}  $D$-string  action, the  $D$-3-brane action is {\it invariant}
under the $SL(2,R)$ transformations of the background fields of 
type IIB theory combined with world-volume vector duality. This conclusion 
 is consistent with  the expectation \jon\ 
that while there are $SL(2)$ multiplets of string (and 5-brane) solutions in type IIB theory, the self-dual 3-brane solution \refs{\hos,\dul} should be unique and related  observation about  the
absence of bound states of  $D$-3-branes was made in  \witt.

The  supersymmetric self-dual 
3-brane solution of type IIB  effective field equations  
\refs{\dul} (extreme limit of  black 3-brane of \hos) 
has non-zero value of $C_4$ field ($C_{mnkl} \sim  \ep_{mnkl} f(x^i) $, 
\  $\del_{[i_1} C_{i_2  i_3 i_4 i_5] }\sim \ep_{i_1 i_2 i_3 i_4 i_5 i_6} \del_{i_6} f$)
and curved metric.
It should be possible to obtain this background 
 as a solution of the 
combined action 
 of type IIB low-energy field theory \efec\  and  $D$-3-brane \ded, 
i.e. it should be supported  by  the  $D$-3-brane source.
 Assuming that all the background
fields except the  metric  $G_{mn}$ and $C_4$  are trivial  
 and that  a consistent choice for the 3-brane  fields 
is  $X^m=x^m, \ X^i=0, \ A_m=0$,  one finds that  
 differentiating \ded\ over $C_{\m\n\l\k}$
produces  a $\d^{(6)}$-function source in the equation for 
$C_{4}$, $\ d^*dC_4 \sim \d^{(6)}(x^i)$.
Since the action \efec\  is  used  under the  prescription that
that the  resulting equations of motion should 
be consistent with self-duality of $F(C_4)$, 
one should  also add the same source to the Bianchi identity
$\ ddC_4 \sim \d^{(6)}(x^i)$. 
The  self-duality equation for  $dC_4$ 
 then holds even in the presence of the 3-brane  source.
 This prescription is in agreement with what follows
  from 
demanding the conformal invariance in the presence 
of 3-brane \doug\ and is also consistent with 
 the interpretation of the  
 solution  \refs{\hos,\dul} as having   both  `electric' and `magnetic' charges.\foot{There should  exist
a more systematic  approach to derivation of the
 full set of equations of motion from an action principle 
based on doubling of the R-R fields (cf.\sss). Then the  original 
 `electric' $C_4$ and its  `magnetic' double $\bar  C_4$ 
 will appear in both the effective type II action and  the 
$D$-brane action  and should automatically  lead to a consistent  set of 
field equations  with `dyonic' sources.
The dyonic nature of  3-brane  is reflected also in the fact 
that the 3-brane source  effectively drops out from 
 the Einstein equations: for  the 3-brane  metric 
$ds^2= f^{-1/2} dx^mdx^m + f^{1/2} dx^idx^i$ 
the $(mn)$ and $(ij)$ components of the Einstein 
equations ($R_{\m\n} \sim (F_5)^2_{\m\n}$)
reduce to $f^{-2} \del^i\del_i f=0$ and 
$f^{-1} \del^i\del_i f=0$ which are satisfied everywhere for $f=1 + Q/r^4$
 (equivalently, that means that the 3-brane source $\d$-function will
be multiplied by a power of $r$ near $r=0$).
In contrast to  the case of  other `elementary'
$p$-brane solutions here the 
 same conclusion applies also 
to the  Einstein equations with raised indices
(as they follow from the effective action).
This may be related to  non-singularity
of the black 3-brane solution  \ght.}
The resulting 3-brane  field configuration  is  then 
the same as in \dul.\foot{It  was  found   in \dul\
that the supersymmetric 3-brane solution 
preserves half of $N=2b,D=10$ supersymmetry
and  thus should be described by  $8+8$ on-shell degrees of freedom 
of $d=4,N=4$ Maxwell supermultiplet $(A_m, X^{IJ}, \l^I)$.
Since 4 other supersymmetries are
spontaneously  broken, they should be realised in a 
non-linear way, so   it was conjectured 
that the resulting action should be of  an unusual  Born-Infeld type \dul. 
This conjecture is indeed confirmed by the  above 
$D$-brane construction of the 3-brane action.}

\newsec{Role of world-volume vector field 
 and higher  dimensional  interpretations of  $D$-branes of type II theories}
Let us draw some lessons from the above discussion 
of $D$-brane actions.
One of the conclusions is  that in the case of type IIB $D$-branes
 the duality 
transformation of the world-volume  vector $A_m$ 
 is closely  related to 
(and, in fact, may  be the origin of)  the  $SL(2,Z)$  symmetry of type IIB theory.
The combined  type IIB + $D$-brane  effective  action 
is covariant under $SL(2,Z)$  provided one also performs
the duality rotation of  $A_m$.
This  is  strongly  analogous to the relation between the
world-sheet 
$d=2$ scalar duality and 
the target space  duality  symmetry of  the string-theory  effective action.
In fact, the world-sheet duality symmetry 
is the reason why the string effective action is 
invariant under $T$-duality. The sum of string effective action  and  fundamental 
 string source action  is $T$-duality covariant provided the  isometric
string source coordinates 
are transformed  simultaneously with the background fields.  
This analogy suggests that if 
the  type IIB superstring theory  could be   interpreted  also 
 as a theory of fundamental supersymmetric  3-branes,  
then  its $SL(2,Z)$ symmetry would be a 
consequence of the  $d=4$ `electro-magnetic'  world-volume 
vector duality. 
 Indeed,  just as the string partition function in an isometric 
background  is $T$-duality invariant
(since the string coordinates are integrated out, their transformation is irrelevant), the  fundamental 3-brane partition function 
in  type  IIB theory background will be invariant under the 
$SL(2,Z)$  transformations \sssl\ (since the vector field $A_m$
is an integration variable, the  vector duality 
transformation $A_m \to \A_m$  which accompanies \sssl\ 
is irrelevant).

The vector duality  plays a different role in the case of  type IIA 
$D$-2-brane: it reveals a hidden $O(1,10)$ 
Lorentz symmetry  making possible to interpret 
the resulting action is a dimensional reduction 
of the  $D=11$ supermembrane  action \refs{\pkt,\schm}.
This  implies a special role played by  the $D$-2-brane
in type IIA theory. 
The above remarks suggest that $D$-3-brane plays analogous special 
role in type IIB theory.

Similarity  with 2-brane case  suggests   that the  dimension 
`associated' with $D$-3-branes is 12.
Indeed,  recall that  the $D$-p-branes 
of type II theories are described 
by  the the  reparametrisation  invariant 
actions depending on the 
fields  $X^\m,\  A_m$  ($\m=0,...,10; \ m=0,...,d-1;   \  d=p+1)$.
As was mentioned already, 
the  number of the physical   degrees of freedom is the same for  all 
 values of 
$d=p+1$: 
$10-d + d-2 =8$, 
where  $10-d$ is  the number of transverse coordinates  $X^i$
remaining, e.g.,  in the static gauge $X^m(x)=x^m$ and 
$d-2$ is the number of transverse modes of a vector in $d$
 with the standard $U(1)$ Maxwell
kinetic term  (leading order  term in  the expansion of \ddw).\foot{Equivalently, one may fix reparametrisation invariance  by choosing 
 $A_m = f_m (x)$  to be   some given  functions but then one  gets $-2$ as  
$U(1)$ ghost contributions.} Suppose we  fix  the $U(1)$ gauge 
but   keep  the reparametrisation invariance
with an  idea  to  reinterpret  the   $D$-p-brane action  as another 
reparametrisation invariant $p$-brane  action.
 Then 
the  number of  `partially  off-shell' 
 bosonic 
degrees of freedom  becomes
  $10+ d-2= 9+p$. This  gives  {\it ten }  for a string,\foot{The critical dimension of the  $D$-string should not, of course,  change 
from its standard (10 or 26) value, in spite of the presence of an extra 
$U(1)$ vector field in its action. 
 Indeed, let us consider  for simplicity the purely bosonic  case.
The fundamental $D$-string is described by a  formal path integral
$Z= \int [dX][dA] \exp [- T \int d^2 x \sqrt {\det (g_{mn} + F_{mn}) }],$ 
where $g_{mn} = \del_m X^\m \del_n X^\m$
is also used  in the definition 
of the measure $[dx][dA]$.  As in the Nambu string case, to make  
this path integral  tractable one  needs to add
some auxiliary fields (like independent 2d metric in Polyakov approach).
Since we are interested  here in  the effect of integrating out $A_m$, 
it is sufficient to introduce just one auxiliary field $V$ as in \bro,
{\it defining} $Z$ as (cf.\bro,\dua) 
$Z= \int [dX][dV][dA] \exp [- \ha T \int d^2 x  \sqrt {\det g_{mn}}
[V (1 + \ha F_{mn} F^{mn})  +  V\inv ]. $
Integrating out $A_m$  we do not get any non-trivial
determinants (the $U(1)$ ghost contribution cancels 
out for the same reason why 
a propagating 2d vector field has zero 
degrees of freedom). We are  then left  with an  action which is
 (semiclassically) equivalent to the Nambu action and thus has the same 
critical dimension 26.} {\it eleven} 
 for the 2-brane,  {\it twelve }  for 3-brane,  etc.  
At the same time, the  amount of on-shell supersymmetry 
 should still remain the same for all $p$-branes and their `higher-dimensional'  versions 
(there are 8 fermionic  physical degrees of freedom as demanded
by unbroken  supersymmetry \pol). 

This indeed is what happens in the case of the  2-brane -- $D=11$  membrane
connection \pkt. 
Since the vector is dual  to a scalar in $d=3$ but to a vector 
 in $d=4$,   in the case of 3-brane  the two extra degrees 
are not 
easily  combined 
with  10 $X^\m$ scalar fields  in a  $D = 12$ Lorentz-invariant way.
Thus if  type IIB superstring theory admits 
a  3-brane reformulation,   its   12-dimensional 
interpretation  should be more subtle than  the 11-dimensional 
one of type IIA theory  implied by the $D$-2-brane -- supermembrane connection:
there should be no  simple $D=12$ Lorentz 
group  relating   all  of the  corresponding  (10+2)  bosonic
degrees of freedom.
 This suggests that the two
extra dimensions  (effectively  represented  by   the `transverse'
part of $A_m$)  may  play only an auxiliary role  in  some   novel 
realisation  of $D=10$ supersymmetry.

Suggestions about possible relation of 
type IIB theory to a 12-dimensional theory were  previously 
made also in 
\refs{\dub,\hull,\vaf}.
It was  conjectured  in \dub\  that there may exist 
a 3-brane with world-sheet signature $(2,2)$\foot{The 4-parameter 
diffeomorphism invariance is, in principle, 
 sufficient to gauge-fix 
 more than one  time-like  direction.  Ref. \dub\ discussed various $p$-branes 
with several time-like coordinates.}
moving in a  $D=12$ space with signature $(2,10)$
 which  reduces upon double-dimensional reduction 
to the usual type IIB superstring  just 
like the  $D=11$ supermembrane  reduces  to the 
type IIA string \duf\ (this  was motivated 
by the fact  that in the case of $(2,10)$ 
signature there exist both Majorana-Weyl spinors
and self-dual tensors in the  supersymmetry algebra).
A picture of 
$(2,2)$ world-volumes  embedded  in  $(2,10)$ space  
 independently  emerged 
 in the context of $N=2$  string theory \ova\ (where 
the `transverse' 8-dimensional space was treated as  an internal one, 
compactified on a special torus). 
The existence  of $D$-string in  type IIB theory 
with an extra vector field on the world sheet  
 was used in \vaf\  to  conjecture (by 
 analogy with  $N=2$ strings  where there is a local $U(1)$ gauge symmetry on the world-sheet)
 that there may exist an `off-shell' extension
of type IIB theory where  two  extra compact  $(1,1)$ dimensions are 
added both to the world-sheet and the target space.  
Type IIB strings would then be recovered by wrapping 
 the $(1,1)$ part of world-volume around the compact $(1,1)$
part of the target space as in \dub.
 A possible 12-dimensional origin 
of $SL(2,Z)$ symmetry of type IIB theory was conjectured  in 
\hull\ (were the signature of a  $D=12$ space-time  
was assumed  to be  the standard one). 
These conjectures may be related to the  systematic approach \kut\ 
to a unified description of different  world-sheet 
string  (and membrane) theories
  as effective target space theories \gri\
of $N=(2,1)$ heterotic string.
 Here   the central role  is played 
by    $(2,2)$
self-dual geometries embedded in  $(2,10)$ space
(with non-compact transverse string coordinates appearing as
zero modes of $N=(2,1)$  string).

Suggestions  of  a  formulation based on 
 $(2,2)$ world-volumes  in $(2,10)$ space-time
 are apparently  different from what was
found above for the $D$-3-brane.
Since   all 8 transverse degrees of freedom
 $X^i$  and $A^\perp_m$  have the same signs of kinetic terms
it seems natural  
  to assign the  standard spatial signature to the extra 
two dimensions.
Also, the $D$-3-brane has standard $(1,3)$ world-volume  signature
(in fact, all $D$-branes have $(1,p)$ signature).
 Finally, 
 one should  not expect to find  the  full 12-dimensional Lorentz (super)symmetry.  
It would be interesting to see if the 
$(1,11)$+$(1,3)$ picture directly implied by the $D$-3-brane description 
can still be  related  to the 
$(2,10)$+$(2,2)$ proposals.\foot{ To relate the 
$(1,3)$ and $(2,2)$ theories one presumably needs to 
perform  a  transformation  
analogous to interchanging the 
left and right  string modes for a time-like circle as in \vaf\
(I am grateful 
to C. Vafa for this  suggestion).
In addition, to relate
$(1,11)$ and $(2,10)$  descriptions 
one  needs  to   `twist'
the vector  field degrees of freedom. 
Twisting  of  4d vector field 
may  be  analogous to  twisting  of 2d  chiral scalars:
in the `doubling'  approach to construction of 
 manifestly dual actions \refs{\tsd,\sss}
the $d=2$ scalar and $d=4$ vector Lagrangians have similar structure
$L_d ({\cal A} ) =  - \ha  \del_0 {\cal A}  {\cal L }  \del {\cal A} 
 + \ha \del {\cal A} {\cal M} \del {\cal A}$, where  ${\cal A}  = (A, \A)$, \
$A$ and $\A$ are $k$-forms (spatial parts of gauge potentials), 
$k = \ha (d-2)$, 
and  ${\cal L} = \pmatrix{  0 & I \cr
(-I)^k  & 0 \cr},$ $ \    
{\cal M} = \pmatrix{  I & 0 \cr
0  & I \cr}.$}

To conclude,  it  is  likely  that the self-dual  Dirichlet 
3-brane  plays a special role in  type IIB superstring theory,  
explaining  its  $SL(2,Z)$ symmetry   
 in terms of  the world-volume `electro-magnetic' duality and 
also pointing towards  an unusual  non Lorentz invariant 
12-dimensional reformulation 
of this theory.  Absence of manifest Lorentz symmetry
is characteristic to models  which describe
chiral $d$-forms  \refs{\mark,\flor}  and also to models
with  doubled numbers 
of  field  variables 
 (but the same number of degrees of freedom)
which are  manifestly duality invariant  \refs{\tsd,\sss}.
It may be  that 12-dimensional theory  is   related to  
such   `doubled' formulation of a self-dual theory.

\bigskip

\noindent {\bf Acknowledgements}

I would like to thank   G. Gibbons, 
C. Schmidhuber and P.K. Townsend for  stimulating discussions.  
I acknowledge  also  the support of PPARC, 
ECC grant SC1$^*$-CT92-0789 and NATO grant CRG 940870.
\vfill\eject
\listrefs
\vfill\eject
\end